\documentclass[aps,prd,twocolumn,nofootinbib,%
               superscriptaddress,preprintnumbers,longbibliography]{revtex4-2}

\usepackage{amsmath,amssymb,amsfonts}
\usepackage{graphicx}
\usepackage{tensor}
\usepackage{bm}
\usepackage{mathrsfs}
\usepackage{mathtools}
\usepackage{xcolor}
\usepackage{siunitx}
\usepackage[colorlinks=true,linkcolor=blue,citecolor=blue,urlcolor=blue]{hyperref}
\usepackage{adjustbox}
\usepackage[normalem]{ulem}
\usepackage{comment}
\newcommand{\dd}{\mathrm{d}}

\newcommand{\scri}{\ensuremath{\mathscr{I}}}

\begin{document}

\title{Source-Driven Tails in Kerr Spacetime:\\
       Nonlinear effects in Late-Time Behavior}

\newcommand{\URI}{\affiliation{Department of Physics, 
    University of Rhode Island, Kingston, RI 02881, USA}}    
\newcommand{\URICCR}{\affiliation{Institute for AI \& Computational Research, 
    University of Rhode Island, Kingston, RI 02881, USA}}    
\newcommand{\UMassDMath}{\affiliation{Department of Mathematics, 
    University of Massachusetts Dartmouth, \\ 285 Old Westport Rd., North Dartmouth, MA 02747, USA}}    
\newcommand{\UMassDPhy}{\affiliation{Department of Physics, 
    University of Massachusetts Dartmouth, \\ 285 Old Westport Rd., North Dartmouth, MA 02747, USA}}    
\newcommand{\UMassDCSCDR}{\affiliation{Center for Scientific Computing \& Data-Science Research, 
    University of Massachusetts Dartmouth, \\ 285 Old Westport Rd., North Dartmouth, MA 02747, USA}}    

\author{Som Dev Bishoyi}
\email{sbishoyi@umassd.edu}
\UMassDMath
\UMassDCSCDR


\author{Subir Sabharwal}
\email{subir@uri.edu}
\URICCR

\author{Gaurav Khanna}
\email{gkhanna@uri.edu}
\URI
\UMassDPhy
\UMassDCSCDR
\URICCR

\date{\today}


\begin{abstract}
  \noindent
  We present the long-duration time-domain simulations 
  of scalar-field tails in Kerr spacetimes driven by
  \emph{outgoing} multipolar sources.  Extending the recent work in 
  the literature from Schwarzschild to rotating black holes, we evolve 
  sources with $\ell'=\{0,1,2,3,4\}$ on backgrounds with dimensionless 
  spin $a/M=\{0.0, 0.8, 1.0\}$ and extract the late-time decay rates of 
  measured modes $\ell\le4$ for a nonlinearity-inspired outgoing source
  with a $1/r^2$ fall-off.  In all cases we find the inverse
  power-law index $p_{\ell\ell'}$ to be larger than the source-free 
  Price law values by one unit, i.e. $p^{\text{sourced}}_{\ell\ell'} = 
  p^{\text{Price}}_{\ell\ell'} + 1$. We also include a power-law index 
  value computation for a similar source-driven gravitational wave 
  case $(\ell,m)=(4,4)$ and confirm closely related results 
  in the recent literature.
\end{abstract}

\maketitle
\tableofcontents
\vspace{-1em}

\section{Introduction\label{sec:intro}}

\subsection{Motivation}
\noindent The monumental detection of gravitational waves (GW) by the LIGO–Virgo–KAGRA network have
promoted the ringdown phase of binary-black-hole mergers to a powerful
laboratory for testing the strong-field regime of general relativity
(GR)~\cite{Abbott2016,Berti2015,Berti2025,Yunes2025}.  While the
early ringdown is governed by a discrete set of quasinormal modes
(QNMs) determined by the linearized Einstein equations, the signal at
late retarded times is known to transition to an inverse power-law
``tail'' first predicted by Price half a century ago~\cite{Price1972a,Price1972b}.  
These tails arise from back-scattering off the effective curvature potential 
surrounding the black hole and fall off as $\Psi_\ell\propto t^{-2\ell-3}$ in Schwarzschild
spacetimes when the initial data have compact support (Here $\ell$ denotes the field 
multipole under study).

With the advent of third-generation ground-based
observatories~\cite{Maggiore2020,Reitze2019} and the space-based
mission LISA~\cite{LISA2024}, the sensitivity to late-time ringdown
will improve considerably, making it essential to understand every
physical effect that can modify the traditional picture.  Recent 
work~\cite{Cardoso2024,Riotto2024a,Ling2025} has 
demonstrated that even within GR, \emph{non-linear} couplings at second
perturbative order produce a \emph{slower} decay for a scalar field on a
Schwarzschild background. In these studies, the lowest order non-linear 
effects are modeled as a source-term in the linear theory and the 
resulting tails are computed. These tails exhibit slower decay when 
compared to Price tails that arise from initial data as opposed to a 
source. Similar conclusions were obtained independently in Ref.~\cite{DeAmicis2024}.  
Because a slower tail may dominate the late-signal, non-linear effects 
must be quantified to avoid systematic biases in spectroscopic analyses.

Rotating (Kerr) black holes are astrophysically generic and introduce
several complications absent in the non-spinning case: $(i)$ mode
mixing between different $\ell$ due to the spheroidal structure of
Kerr QNMs; $(ii)$ frame dragging, which modifies the effective
potential that sources back-scattering; and $(iii)$ an additional
near-horizon, near-extremal (NHEK) region where perturbations can
linger, affecting the late-time field seen by distant
observers~\cite{nek-sign,nek-sign2,nek-sign3,nek-sign4}.  
Although linear tails in Kerr have been studied extensively both
analytically and numerically~\cite{Tiglio:2007jp,Gleiser:2007ti,Zenginoglu:2009hd,Burko:2007ju,Zenginoglu:2012us,Racz:2011qu,Burko:2013bra,Angelopoulos:2021cpg}, the non-linear problem remains largely unexplored. The generalization 
of Price law to sub-extremal Kerr spacetimes for the case of compact 
initial data located with no support on the horizon is given 
by~\cite{Zenginoglu:2012us,Burko:2013bra,Angelopoulos:2021cpg}
\footnote{Sometimes referred to as the ``ZKB formula'' since it 
was first presented in Ref.~\cite{Zenginoglu:2012us}.}:  
\begin{equation}  \label{eqn:rates1}\quad
p_{\ell\ell'} = \left\{ \begin{array}{ll}
-(\ell'+\ell + 3) & \mathrm{for}\quad \ell'=0, 1 \\
-(\ell'+\ell + 1) & \mathrm{otherwise}\quad \end{array}\right.\\
\end{equation}
along $r={\rm const}$ and at $\mathscr{H}^+$, and
\begin{equation} \label{eqn:rates2}\quad
p_{\ell\ell'}^{\scri^+} = \left\{ \begin{array}{ll}
-\ell' & \mathrm{for}\quad \ell\leq \ell'-2  \\
-(\ell+2) & \mathrm{for} \quad \ell\geq \ell' \end{array}\right.
\end{equation}
along null infinity $\mathscr{I}^+$, where $\ell'$ refers to the initial field 
multipole, and $\ell$ is the multipole being observed. These rates refer to 
axisymmetric multipoles, and were obtained by carefully studying the 
inter-mode coupling effects that are present in the Kerr spacetime due 
to the lack of spherical symmetry. The non-axisymmetric results do not 
change, except for fact that both $\ell'$ and $\ell$ are bounded from below by the 
value of $m$~\cite{Burko:2010zj}. The extremal case has a similar behavior with 
the exception that the horizon rates match the null infinity rates due to the 
presence of a conformal symmetry~\cite{symmetry}. 

The primary aim of this work is therefore to \emph{numerically}
generalize the findings of Refs.~\cite{Cardoso2024,Riotto2024a,Kehrberger2025} to Kerr 
spacetimes and to map out the dependence of the late-time power-law 
index $p_{\ell\ell'}$ on both the source multipole $\ell'$ and the observed $\ell$ mode and zero initial data.  
We focus on scalar fields as a computationally economical proxy for the 
gravitational case. However, we also include a result for the lowest-order 
non-linear effect in the gravitational wave case. Our study is the first to 
perform evolutions of non-linearity inspired, source-driven simulations of 
mode-coupled scalar tails for long simulation times $t/M,u/M\gtrsim10^3$ 
using a modern GPU-accelerated WENO code.  The results serve as a crucial 
stepping stone toward full gravitational-wave predictions and have direct 
implications for template accuracy in upcoming high-signal-to-noise 
detections.

\subsection{Key results}
\noindent Our main findings for the case of Kerr tails driven by a
nonlinearity-inspired source with multipole $\ell'$ appear below:

\begin{enumerate}
\item For all spins (including extremal) the numerical power-law 
      tails match the general prediction of Ref.~\cite{Cardoso2024}
      for both  $r={\rm const}$ timelike worldline and along null infinity 
      $\mathscr{I}^+$. More specifically, the power-law follows the values in 
      Eqns.~\ref{eqn:rates1} and~\ref{eqn:rates2} simply increased by one unit. 
\item As argued by Ref.~\cite{Cardoso2024,Riotto2024a} the lowest order 
       effect of nonlinearities should appear in the $\ell=m=4$ gravitational 
       wave mode from the squared $\ell'=m'=2$ mode acting as the source. Our 
       results agree with Ref.~\cite{Cardoso2024} i.e., $t^{-10}$ on a 
       $r={\rm const}$ time-like worldline. Note that Ref.~\cite{Cardoso2024} 
       and Ref.~\cite{Riotto2024a} results appear to be in tension since 
       they make different claims on the power-law index, however the 
       difference can reconciled since their results use different gauge 
       choices.
\end{enumerate}
\noindent In summary, 
\begin{equation}
p^{\text{sourced}}_{\ell\ell'} = p^{\text{Price}}_{\ell\ell'} + 1 
\label{eq:sourced_shift}
\end{equation}
where $p^{\text{Price}}_{\ell\ell'}$ refers the well-known Price law and its 
generalization to Kerr spacetime with no source and initial data $\ell'$ as 
given in Eqns.~\ref{eqn:rates1} and~\ref{eqn:rates2}. Recall that the meaning 
of $\ell'$ is different in the different terms of Eqn.~\ref{eq:sourced_shift}. 
On the left-hand-side of the equation, $\ell'$ labels the multipole value of 
the source term, whereas on the right-hand-side it is the multipole value of 
the initial data.

\subsection{Organization}
\noindent Section~\ref{sec:background} reviews linear and non-linear tail theory
in Kerr. Section~\ref{sec:numerics} details our numerical implementation, 
including the evolution scheme, boundary conditions, parallelization strategy, 
and code-validation tests. Section~\ref{sec:results} presents our power-law 
index extractions, and a comprehensive set of tables for $\ell',\ell\le4$ across a 
sub-extremal spin value.  The extremal limit is treated separately in Sec.~\ref{sec:extremal}.  
We conclude in Sec.~\ref{sec:concl} with an outlook toward future calculations
and observable consequences.  

\section{Background\label{sec:background}}
In this section, we briefly describe the Teukolsky equation and the coordinate 
systems used to derive the evolution equations that are solved numerically. We also 
describe how an outgoing pulse whose amplitude falls off radially can be used as 
a model for a source term inspired from non-linearities.

\subsection{Homogeneous Teukolsky equation}
\noindent We consider perturbations $\Psi(t,r,\theta, \phi)$ propagating on a fixed 
Kerr geometry of mass $M$ and angular momentum $J=aM$. The dynamics of scalar, vector 
and tensor field perturbations in the spacetime of Kerr black holes is governed by the 
Teukolsky equation:
\begin{eqnarray}
\label{eq:teuk0}
&&
-\left[\frac{(r^2 + a^2)^2 }{\Delta}-a^2\sin^2\theta\right]
         \partial_{tt}\Psi
-\frac{4 M a r}{\Delta}
         \partial_{t\phi}\Psi \nonumber \\
&&- 2s\left[r-\frac{M(r^2-a^2)}{\Delta}+ia\cos\theta\right]
         \partial_t\Psi\nonumber\\  
&&
+\,\Delta^{-s}\partial_r\left(\Delta^{s+1}\partial_r\Psi\right)
+\frac{1}{\sin\theta}\partial_\theta
\left(\sin\theta\partial_\theta\Psi\right)+\nonumber\\
&& \left[\frac{1}{\sin^2\theta}-\frac{a^2}{\Delta}\right] 
\partial_{\phi\phi}\Psi +\, 2s \left[\frac{a (r-M)}{\Delta} 
+ \frac{i \cos\theta}{\sin^2\theta}\right] \partial_\phi\Psi  \nonumber\\
&&- \left(s^2 \cot^2\theta - s \right) \Psi = 
0,
\end{eqnarray}
where $M$ is the mass of the Kerr black hole, $a$ its angular momentum per unit 
mass, $\Delta = r^2 - 2 M r + a^2$. The $s = 0$ version of this equation describes 
scalar fields. When $s = -2$  the Teukolsky master variable describes outgoing 
gravitational radiation and is related to the perturbed Weyl Scalar $\Psi_{4}$ 
by $\Psi = \rho^{-4}\Psi_{4}$ where $\rho = -1/(r-ia\cos{\theta})$. Unless stated otherwise, we use $\Psi$ to denote the first-order perturbation $\Psi^{(1)}$ about the background value, in accordance with standard practice in the literature.\\

Due to the presence of coordinate singularities in the metric at $\Delta=0$, 
we switch to a better suited coordinate system known as ingoing Kerr coordinates. 
These coordinates are able to smoothly penetrate the horizon of a black hole. 
These new coordinates $(\tilde{t},r,\theta, \tilde{\phi})$ are related to the 
Boyer-Lindquist coordinates by
\begin{align}\label{eq:ingngKerr}
    \tilde{\phi} = \phi + \int \frac{a}{\Delta}dr
    \\
    \tilde{t} = t-r+r_*
\end{align}
where the tortoise radial coordinate $r_* = \int (r^2+a^2)\Delta^{-1}dr$. At late 
times, the $\tilde{t}$ variable essentially becomes the null variable $v = t + r_*$ at the horizon.
The next step that goes into the setup of our coordinate system is hyperboloidal 
compactification as developed by Zengino\v{g}lu~\cite{cauchy}. To do this, we define 
a compactified coordinate system $(\tau,\rho,\theta,{\varphi})$ by
\begin{equation}
\tau = \tilde{t} - {r}^2/({r}+S) + 4 \ln [S/({r}+S)]
\end{equation}
and
\begin{equation}
\rho ={r}/[1+{r}/S]
\end{equation}
where a free parameter $S$ controls both the domain and also the foliation. Note 
that $\rho\in [0,S)$ maps $r\in [0,\infty )$ and is therefore a one-to-one compactifying 
coordinate. A Penrose diagram of the slices defined by these coordinates in the Kerr 
spacetime context can be found in Ref.~\cite{cauchy}. We do not show the final form of 
the equation in these compactified coordinates because of the lengthy nature of the 
expression and the fact that it is not particularly illuminating. In symbolic form, 
it can be written as
\begin{align}\label{eq:teuksymb}
    A^{\tau\tau}\,\partial_{\tau}^{2}\psi
+ A^{\tau\rho}\,\partial_{\tau}\partial_{\rho}\psi
+ A^{\rho\rho}\,\partial_{\rho}^{2}\psi
+ A^{\theta\theta}\,\partial_{\theta}^{2}\psi \nonumber\\
+ B^{\tau}\,\partial_{\tau}\psi
+ B^{\rho}\,\partial_{\rho}\psi
+ B^{\theta}\,\partial_{\theta}\psi
+ C\psi = 0,
\end{align}
Additional details on writing Eq.~\eqref{eq:teuksymb} in first-order form and the particular 
choice of auxiliary variables used can be found in \cite{weno}. 

\subsection{Non-linearity inspired source}
\noindent To understand the behavior of non-linear effects in perturbation theory, we consider 
the second order quantity in the perturbative expansion of the unperturbed scalar $\Psi$, which we 
write as $\Psi^{(2)}$. Campanelli and Lousto first showed that the evolution of $\Psi^{(2)}$ 
is also governed by the Teukolsky equation, with a source term that depends on all components of 
the first order metric perturbation $h_{\mu\nu}$ \cite{Campanelli:1998jv}. The equation of motion 
for $\Psi^{(2)}$ is
\begin{eqnarray}
    \mathcal{T}\Psi^{(2)} = \mathcal{S}[h^{(1)}_{\mu\nu}]
\end{eqnarray}
where $\mathcal{T}$ is the Teukolsky operator and $\mathcal{S}$ is the source term which depends 
on the first order metric perturbation $h^{(1)}_{\mu\nu}$.

Asymptotic behavior of source terms in the second-order Teukolsky equation decay as $\mathcal{O}(1)/r^2$ 
at infinity (See Sec. 5 in \cite{Spiers:2023cip}). This source term can be written as 
\begin{equation}
    \mathcal{S}[h^{(1)}_{\mu\nu}] = \frac{F(u,\theta, \phi)}{r^2} + \mathcal{O}(r^{-3})
\end{equation}
where $u = t-r_*$ is the retarded time coordinate. As a simpler and computationally light-weight 
approximation, we choose a Gaussian source moving radially outward with the appropriate falloff 
to model an outgoing first-order perturbation. The resulting non-linearity inspired source becomes 
\begin{equation} \label{eq:nonlinsrc}
    \mathcal{S}(u,\theta,\phi) \sim  \frac{G(u - u_{0})}{r^{2}} Y_{\ell'm'}(\theta, \phi)
\end{equation}
where $G$ is a standard Gaussian function.
\section{Numerical Implementation\label{sec:numerics}}
\noindent Our numerical implementation scheme entails re-writing the second 
order partial differential equation (PDE) above in terms of two coupled first-order 
differential equations. We solve this system using a high-order weighted 
essentially non-oscillatory (WENO) finite-difference scheme with explicit 
Shu-Osher time-stepping. Details may be found in our previous work~\cite{weno}. 
In the Schwarzschild case, we set $S=18.0$, while in the Kerr cases we 
choose $S=18.4$ for $a/M=0.8$ and $S=19.0$ for $a/M=1.0$. 
The source is a Gaussian of unit width in $u$ with a radial fall-off of $1/r^2$ which 
essentially is the $\beta=2$ case in Ref.~\cite{Cardoso2024}. The initial data is set to 
zero, so we only obtain outgoing source-driven tails in this work and not those driven 
by initial data. The angular distribution of the source is characterized by the $\ell'$ 
multipole. We use high-precision floating-point arithmetic, so that we can track decaying 
values for long durations accurately. Specifically, we use 128-bit quadruple-precision, 
which is particularly important for accurately studying the $\ell=4$ multipole mode.

Finally, to complete these long duration, high-accuracy and high-precision 
computations in a reasonable time-frame we make extensive use GPGPU-based 
parallel computing. For additional details on implementation of such intensive 
computations on a parallel GPU architecture, we refer the reader to our earlier 
work on the subject~\cite{weno}. Note that these simulations take significant 
computational resources to run. For this reason, we focused our efforts on 
a single value of $a/M$ for the sub-extremal ($0< a/M < 1$)   case. 

\subsection{Simulation catalog}
\noindent We carried out $3\times5=15$ primary simulations covering source 
$\ell'=\{0,1,2,3,4\}$ and spins $a/M=\{0.0,0.8,1.0\}$. Typical resolution used was 
$(N_{\rho},N_\theta)=(32000,\pi/128)$ with a Courant factor of $0.64$. Each computation 
took approximately a day on an Nvidia V100 GPU with an IBM POWER9 host. 

It is worth repeating here that the evolutions include a nonlinearity-inspired 
source with multipole $\ell'$ and zero initial data. This is different from the 
classic Price law cases wherein the initial data has a multipole $\ell'$ and there 
is no source. 

\subsection{Power-law extraction}
\noindent For each $(\ell',\ell)$ we compute the instantaneous log-slope (so-called ``LPI'' for 
local power-law index). For example, at null infinity this quantity takes the form
\begin{equation}\label{eq:LPI}
  p_{\ell\ell'}(u)
  \;=\;
  \frac{\dd\ln|\Psi_\ell|}{\dd\ln u},
\end{equation}
and is computed using a centered 4-point finite-difference stencil in $\ln u$.  We 
then fit $p_{\ell\ell'}(u)$ over the last few $100M$ to a constant plus a term $\propto u^{-1}$, 
consistent with theoretical expectations for the next-to-leading order tail. The 
extrapolated constant is quoted as the power-law index; statistical fit uncertainties 
and the PDE solver truncation error dominate the error budget.

\section{Results\label{sec:results}}

\subsection{Results: Sub-Extremal Kerr\label{sec:results}}
\begin{figure}[t]
  \centering
  \includegraphics[width=0.5\textwidth]{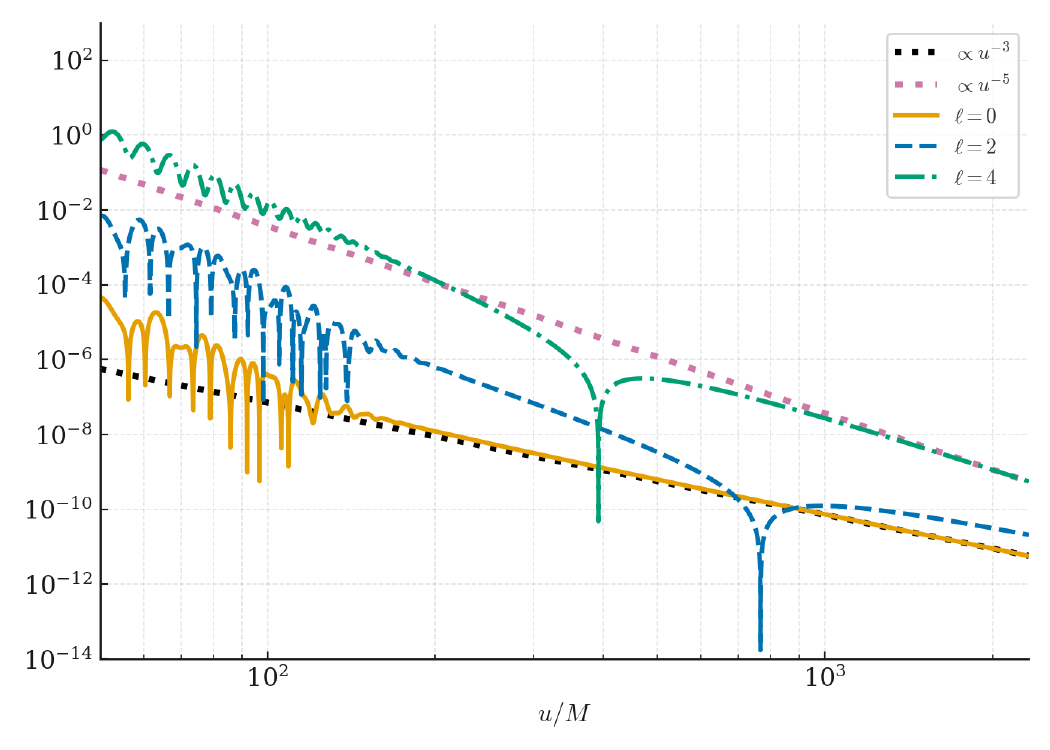}
  \caption{Log-log plot of different angular projections $\ell=0,2,4$ of the scalar field $\Psi^{(2)}$ vs.\ $u$ at $\scri^+$ for $\ell'=4$ source for a Kerr black hole with $a/M=0.8$.}
  \label{fig:example_a08}
\end{figure}

\noindent Figure~\ref{fig:example_a08} shows the late-time decay of  $\Psi^{(2)}$ sourced by 
$\ell'=4$ on a background with $a/M=0.8$. The slopes converge to $\{-3,-3,-5\}$ 
for $\ell=\{0,2,4\}$ in agreement with Eqn.~\ref{eq:sourced_shift}.  Similar behavior 
occurs for all other mode pairings.
Tables~\ref{tab:LPI_a0}–\ref{tab:LPI_a08} list the extracted 
for each spin. The Schwarzschild results are in clear agreement with 
Ref.~\cite{Cardoso2024}. For all spinning cases the numerical power-law 
tails match the general prediction of Ref.~\cite{Cardoso2024} for both  
$r={\rm const}$ timelike worldline and along null infinity $\mathscr{I}^+$ 
in the sense that the power-law follows the values in Eqns.~\ref{eqn:rates1} 
and~\ref{eqn:rates2}, simply increased by one unit.
\begin{table}[b]
  \caption{Inverse-power indices $-p_{\ell\ell'}$ for Schwarzschild ($a=0$) at the locations: $r$=const. and $\scri^+$.}
  \label{tab:LPI_a0}
 \setlength{\tabcolsep}{0.5pt}
 \begin{adjustbox}{width=\columnwidth}
  \begin{ruledtabular}
  \begin{tabular}{c|ccccc}
     & \multicolumn{5}{c}{$\ell$}\\
     $\ell'$ & 0 & 1 & 2 & 3 & 4\\ \hline
     0 & 1.999, 1.004& --& --& --& --\\
     1 & --& 4.048, 2.014& --& --& --\\
     2 & --& --& 5.989, 3.087& --& --\\
     3 & --& --& --& 7.983, 3.999& --\\
     4 & --& --& --& --& 9.813, 5.042\end{tabular}
  \end{ruledtabular}
\end{adjustbox}
\end{table}
\begin{table}[b]
  \caption{Inverse-power indices $-p_{\ell\ell'}$ for Kerr ($a=0.8$) at the locations: $r$=const. and $\scri^+$.}
  \label{tab:LPI_a08}
 \setlength{\tabcolsep}{0.5pt}
 \begin{adjustbox}{width=\columnwidth}
  \begin{ruledtabular}
  \begin{tabular}{c|ccccc}
     & \multicolumn{5}{c}{$\ell$}\\
     $\ell'$ & 0 & 1 & 2 & 3 & 4\\ \hline
     0 & 1.998, 1.003& --& 3.994, 2.993& --& 6.110, 4.800\\
     1 & --& 4.050, 2.011& --& 5.992, 4.011& --\\
     2 & 2.001, 1.008& -- & 3.901, 2.997& --& 6.210, 5.010\\
     3 & --& 3.998, 2.020& --& 6.020, 3.952& --\\
     4 & 3.920, 3.036& --& 6.206, 2.861& --& 7.752, 4.820\end{tabular}
  \end{ruledtabular}
 \end{adjustbox}  
\end{table}

These results are somewhat counterintuitive. One may reason that with 
a $\ell'$ source, a $\ell=\ell'$ mode would develop that would behave  
just as in Schwarzchild spacetime~\cite{Cardoso2024}. Additionally, 
through mode-coupling in Kerr, infinitely many $\ell\neq\ell'$ modes would 
get ``excited''~\cite{Burko:2013bra}. Now, these modes are not source-driven 
since the source-term is non-zero only for the $\ell'$ case. Therefore, it 
is reasonable to argue that those $\ell\neq\ell'$ modes would evolve just as 
they do in the source-free cases, i.e. simply follow Price law in Kerr as in 
Eqns.~\ref{eqn:rates1} and~\ref{eqn:rates2}. However, that is {\em not} 
what we observe at all. Our results show that {\em all} $\ell$ modes 
follow Eqn.~\ref{eq:sourced_shift} and decay slower than what one may have argued 
through intuitive reasoning. Further investigation is needed to fully understand 
this rather intriguing result. 

\subsection{Results: Extremal Limit\label{sec:extremal}}
\noindent We can take $a=M$ using our numerical approach and study the late-time 
behavior of the scalar field multipoles. Table~\ref{tab:extremalLPI} presents 
the results. One expects the $r={\rm const}$ and null infinity $\mathscr{I}^+$ tails 
to behave the same as the sub-extremal case, however the horizon tails would be 
different. For this reason we present the horizon and $\mathscr{I}^+$ rates only. 
\begin{table}[t]
  \caption{Inverse-power indices $-p_{\ell\ell'}$ for the extremal Kerr at the locations: $\mathscr{H}^+$ and $\scri^+$.}
  \label{tab:extremalLPI}
 \setlength{\tabcolsep}{0.5pt}
 \begin{adjustbox}{width=\columnwidth}
  \begin{ruledtabular}
  \begin{tabular}{c|ccccc}
     & \multicolumn{5}{c}{$\ell$}\\
     $\ell'$ & 0 & 1 & 2 & 3 & 4\\ \hline
       0 & 2.005, 0.993& --& 4.009, 3.005& --& 6.034, 5.008\\
       1 & --& 3.002, 2.000& --& 4.970, 4.035& --\\
       2 & 1.895, 1.022& --& 3.970, 3.020& --& 5.919, 5.018\\
       3 & --& 2.992, 2.007& --& 4.980, 4.199& --\\
       4 & 3.854, 3.047& --& 3.920, 2.998& --& 6.161, 4.822\end{tabular}
  \end{ruledtabular}
  \end{adjustbox}
\end{table}
It is interesting to note that the horizon rates here do not appear to match the null infinity 
rates as one may expect due to the conformal symmetry~\cite{symmetry} in extremal spacetimes. 
In fact, they appear to simply match the source-free, initial data driven results, i.e. follow 
the Price law equivalent for extremal Kerr~\cite{Burko2023}. We speculate that it is due to a 
breakdown of this conformal symmetry between $\mathscr{H}^+$ and $\scri^+$ because of the 
presence of an outgoing source-term. Further investigation is needed to fully understand this result. 

It is worth noting that the type of source~\cite{Cardoso2024} we have studied in this work 
does not have any impact on the Aretakis charge values on the horizon. Additionally, as 
shown in \cite{Cardoso2024} only an \textit{outgoing} source changes the power-law decay 
rates from Price tails, thus minimizing an opportunity for the source to leave any footprint 
on the horizon. Thus, the Aretakis charges will continue to arise through the imposition 
of non-trivial initial data on the horizon -- a case that we have studied elsewhere~\cite{Burko2023} 
and not in this work. 

\subsection{Results: Gravitational wave case\label{sec:gravity}}
\noindent Now we turn to the lowest-order gravitational case with non-linear 
effects that has received some attention in the recent literature. 

As argued in Ref.~\cite{Cardoso2024,Riotto2024a} the lowest order 
effect of non-linearities should appear in the $\ell=m=4$ gravitational 
wave mode of $\Psi^{(2)}$ from the squared $\ell'=m'=2$ mode of $\Psi^{(1)}$ as source. Ref.~\cite{Cardoso2024} 
results suggest that the $r={\rm const}$ rates should be $t^{-10}$, while the 
Price tails formula would suggest a tail of $t^{-11}$. 

\begin{figure}[t]
  \centering
  \includegraphics[width=0.5\textwidth]{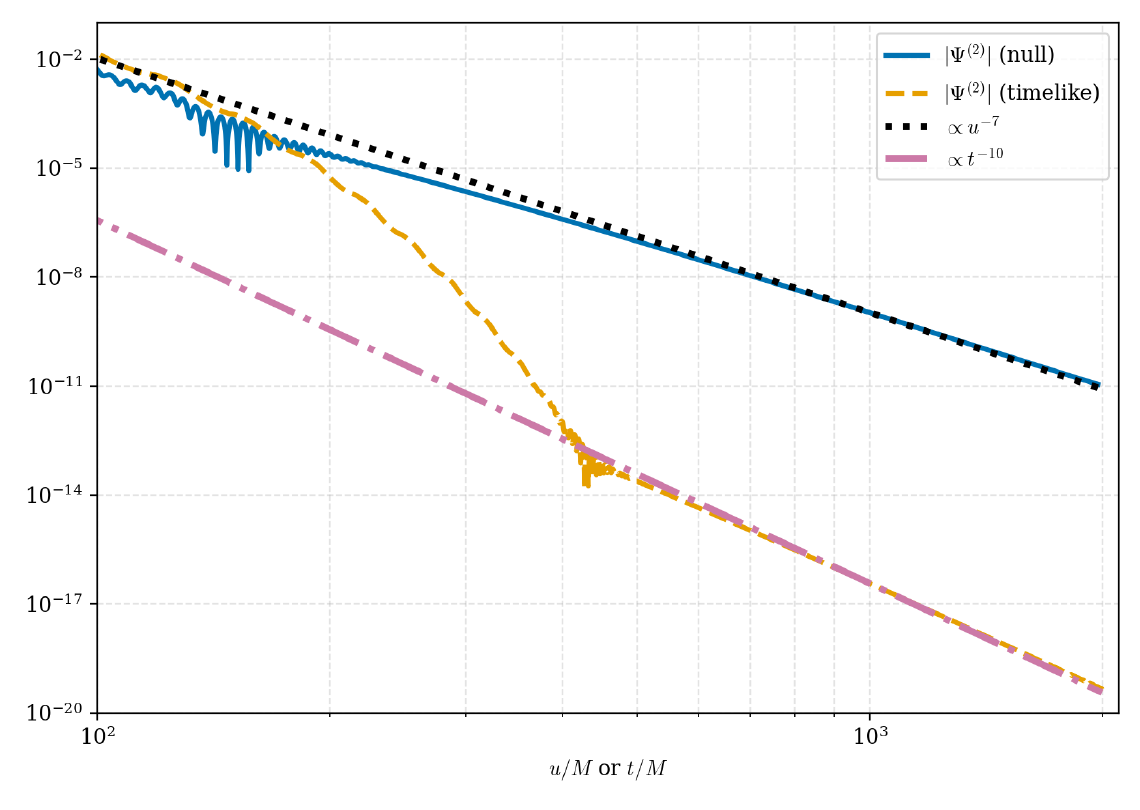}
  \caption{Log–log plot of the $(\ell,m)=(4,4)$ angular projection of $|\Psi^{(2)}|$ 
  vs.\ $t, u$ with $(\ell',m')=(2,2)$ squared source and $a/M=0.8$.} 
  \label{fig:example_gw}
\end{figure}

Our results depicted in Fig.~\ref{fig:example_gw} confirm those results   
i.e. $t^{-10}$ on a $r={\rm const}$ timelike worldline for $|\Psi^{(2)}|$. 
At $\scri^+$ we obtain a late-time behavior of $u^{-7}$ which is one power 
slower than the source-free case~\cite{Racz:2019}. 
%
Our results have been consistent throughout in the general sense 
that the rates with the nonlinearity-inspired source decay  
slower than Price law by a single power.
\section{Discussion and Conclusions\label{sec:concl}}
\noindent In this work we evolve scalar fields with outgoing sources with 
$\ell'=\{0,1,2,3,4\}$ on black hole backgrounds with dimensionless spin
$a/M=\{0.0, 0.8, 1.0\}$ and extract the late-time decay rates of measured
modes $\ell\le4$.  In all cases we find the inverse power-law index to be 
larger than the source-free Price law values by one unit, i.e. 
$p^{\text{sourced}}_{\ell\ell'} = p^{\text{Price}}_{\ell\ell'} + 1$. This 
result is counterintuitive given the common understanding of mode-coupling 
in Kerr spacetime. More specifically, given that the source is only non-zero 
for multipole $\ell'$, one would expect the tails for the $\ell\neq\ell'$ 
cases to not change.

We also study a power-law index value for the gravitational wave case $(\ell,m)=(4,4)$ 
driven by a $(\ell',m')=(2,2)$ squared source, and confirm related results 
in the recent literature. 

The form of the outgoing source is inspired by the lowest order nonlinearity 
that is envisioned to appear in second-order perturbation theory. Our results 
are interesting because they suggest that with the inclusion of non-linearity, 
the tails decay slower and that may increase their chances of being observed 
by the next generation of gravitational wave observatories. In fact, early full 
numerical relativity simulations~\cite{DeAmicis2024} offer some compelling 
evidence that indeed, the power-law decays are much slower than those 
predicted by Price law.

Future natural extensions of this work would be in two directions: $(i)$ more 
extensive study of the gravitational wave case, and $(ii)$ a broader study of the 
nonlinear terms that may impact the expected outcomes from linear theory. 

To conclude, the late-time behavior of a perturbed Kerr black hole is more 
intricate than linear theory foretold, yet its leading imprint can still be 
captured by a remarkably simple power law.  We hope the present work paves the
way for a comprehensive, nonlinear understanding of black-hole ringdown in 
the era of precision gravitational-wave astronomy.

\begin{acknowledgments}
\noindent G.K. acknowledges support from NSF Grants No. PHY-2307236 and DMS-2309609.
S.B. acknowledges support of NSF grants PHY-2110496 and DMS-2309609. All 
computations were performed on the UMass-URI UNITY HPC/AI cluster at the 
Massachusetts Green High-Performance Computing Center (MGHPCC). We also 
acknowledge the use of OpenAI's ChatGPT for assistance with summarizing some 
background material in an initial draft.
\end{acknowledgments}


\bibliographystyle{apsrev4-2}





\end{document}